\documentclass{sigchi}%
\usepackage{balance}
\usepackage{graphics}
\usepackage[T1]{fontenc}
\usepackage{txfonts}%
\usepackage{mathptmx}%
\usepackage{xcolor}%
\usepackage{booktabs}%
\usepackage{textcomp}%
\usepackage{microtype}
\usepackage[utf8]{inputenc}%
\usepackage[all=normal,wordspacing,floats,paragraphs]{savetrees}%
\usepackage{makecell}%
\usepackage{multirow}
\usepackage{paralist}%
\usepackage{tabularx}%
\usepackage{subcaption}%
\usepackage{comment}%
\usepackage[pdflang={en-US},pdftex]{hyperref}
\toappear{\scriptsize Permission to make digital or hard copies of all or part of this work for personal or classroom use is granted without fee provided that copies are not made or distributed for profit or commercial advantage and that copies bear this notice and the full citation on the first page. Copyrights for components of this work owned by others than ACM must be honored. Abstracting with credit is permitted. To copy otherwise, or republish, to post on servers or to redistribute to lists, requires prior specific permission and/or a fee. Request permissions from permissions@acm.org. \\
{\emph{CHI '20, April 25--30, 2020, Honolulu, HI, USA.} } \\
Copyright is held by the owner/author(s). Publication rights licensed to ACM. \\
ACM ISBN 978-1-4503-6708-0/20/04\ ...\$15.00.\\
http://dx.doi.org/10.1145/3313831.3376677}

\newcommand\todo[1]{{{\color{black}#1}}}%
\newcommand\rev[1]{{{\color{black}#1}}}%
\newcommand\countandprint[1]{\stepcounter{#1}\arabic{#1}.}%

\newcolumntype{b}{X}
\newcolumntype{s}{>{\hsize=.7\hsize}X}

\def\plaintitle{Creativity on Paid Crowdsourcing Platforms}
\def\plainauthor{Jonas Oppenlaender, Kristy Milland, Aku Visuri, Panos Ipeirotis, Simo Hosio}
\def\plainkeywords{Creativity;~crowdsourcing;~creative~work;~creativity~tests; creativity support tools.}

\makeatletter
\def\url@leostyle{%
  \@ifundefined{selectfont}{
    \def\UrlFont{\sf}
  }{
    \def\UrlFont{\small\bf\ttfamily}
  }}
\makeatother
\def\pprw{8.5in}
\def\pprh{11in}

\definecolor{linkColor}{RGB}{6,125,233}
\hypersetup{%
  pdftitle={\plaintitle},
  pdfauthor={\plainauthor},
  pdfkeywords={\plainkeywords},
  pdfdisplaydoctitle=true, 
  bookmarksnumbered,
  pdfstartview={FitH},
  colorlinks,
  citecolor=black,
  filecolor=black,
  linkcolor=black,
  urlcolor=linkColor,
  breaklinks=true,
  hypertexnames=false}

\begin{document}

\title{\plaintitle}

\numberofauthors{1}
\author{%
  \alignauthor{Jonas Oppenlaender\textsuperscript{1}, Kristy Milland\textsuperscript{2}, Aku Visuri\textsuperscript{1}, Panos Ipeirotis\textsuperscript{3}, Simo Hosio\textsuperscript{1}\\
    \affaddr{\textsuperscript{1}Center for Ubiquitous Computing, University of Oulu, Finland}\\
    \affaddr{\textsuperscript{2}Faculty of Law, University of Toronto, Canada}\\
    \affaddr{\textsuperscript{3}Stern School of Business, New York University, New York, NY, USA}\\
    \email{\textsuperscript{1}\{firstname.lastname\}@oulu.fi},
    \email{\textsuperscript{2}\{firstname.lastname\}@utoronto.ca},
    \email{\textsuperscript{3}\{firstname\}@stern.nyu.edu}
  }\\
}

\maketitle

\begin{abstract}
Crowdsourcing platforms are increasingly being harnessed for creative work. The platforms' potential for creative work is clearly identified, but the workers' perspectives on such work have not been extensively documented. In this paper, we uncover what the workers have to say about creative work on paid crowdsourcing platforms. Through a quantitative and qualitative analysis of a questionnaire launched on two different crowdsourcing platforms, our results revealed clear differences between the workers on the platforms in both preferences and prior experience with creative work.~We identify common pitfalls with creative work on crowdsourcing platforms, provide recommendations for requesters of creative work, and discuss the meaning of our findings within the broader scope of creativity-oriented research. To the best of our knowledge, we contribute the first extensive worker-oriented study of creative work on paid crowdsourcing platforms.
\end{abstract}



\section{Introduction}%


Crowdsourcing (CS) on platforms such as Amazon Mechanical Turk (MTurk) and Prolific is increasingly being harnessed by industry and academia in a variety of use cases~\cite{dennis2003,p453-kittur.pdf}.
For researchers, these platforms offer a convenient and
flexible means for collecting survey data and conducting online human-subject experiments~\cite{Gadiraju2-1wgg9voexzjhw3.pdf,jdm10630a.pdf}.
Each year, top venues in Human-Computer Interaction (HCI) and design (e.g., CHI, CSCW, DIS) publish a variety of articles with paid crowd workers as the sole source of participants.
A growing body of research uses paid CS platforms for creativity-oriented research~\cite{2019_chi-paper.pdf,p1235-frich.pdf,p22-kittur.pdf}.

Creativity is considered a grand challenge in HCI~\cite{Shneiderman2009}, and the use of convenient platforms for participant recruitment and eliciting ideas is easy to sympathize with.
\rev{Yet there is a clear gap in the literature which predominantly investigates CS from the perspective of the requester of work.}
Given that empirically based contributions are prevalent in creativity studies in HCI~\cite{p1235-frich.pdf},
it is imperative to develop an understanding of how all stakeholders of creative work perceive creativity.
Insights into the workers' perspective are rare but important, as they may inform the design of studies with higher validity and 
the design of online tools that rely on \textit{crowd-powered creativity}.

In this paper, we focus on two commonly used CS platforms that both compensate crowd workers for completing tasks and participating in online surveys: Prolific and Amazon Mechanical Turk.
We conducted a 
{questionnaire study} 
on these platforms,
focusing on the workers' attitudes and preferences concerning creative work. 
Among other findings, our analysis of responses from 215~workers reveals clear differences between the workers of the two platforms in both preferences and prior exposure to creative work. The key contributions
of this paper
are:%
\begin{compactitem}%
\itemsep0.7mm
\item an in-depth analysis of the worker preferences concerning creative work on Prolific and MTurk,%
\item clear evidence for the nonna\"{i}vet\'{e} of crowd workers in regard to commonly used creativity tests, and%
\item a presentation of five crowd worker archetypes, based on different perceptions and attitudes towards creative work.
\end{compactitem}%
To the best of our knowledge, our work contributes the first extensive worker-oriented qualitative study of creative work on the two commonly used paid crowdsourcing platforms.

Crowdsourcing platforms are excellent sources of participants for creativity-oriented research. \todo{However,} with this paper we wish to raise awareness of some of the shortcomings of the current research and practice in using these platforms for creative work. 
We highlight issues that may affect the validity of crowdsourcing studies and provide recommendations for requesters of creative work on crowdsourcing platforms.
Together with the worker archetypes, our findings help researchers who wish to harness the inherent convenience and power of crowdsourcing platforms in creativity-oriented studies and solutions.%

\begin{table*}%
\centering%
\begin{tabularx}{\textwidth}{
    >{\hsize=.045\hsize}X
    >{\hsize=.006\hsize}X
    >{\hsize=.23\hsize}X
    >{\hsize=.295\hsize}X
    l
    X
}%
    &
    & {\small \textit{Task}}
    & {\small \textit{Task type}}
    & {\small \textit{Examples}}
    & {\small \textit{Examples of task instructions}} \\
    \midrule
    \multirow{3}{*}{\small
        \rotatebox[origin=c]{90}{
            \makecell[c]{Creative\\tasks}
        }
    }
    &
    \small a)
    & \small Ideation
    & \small Divergent thinking
    & \small \cite{IdeationAtScale-cscw2015.pdf,p1245-yu.pdf}
    & \small \textit{``Come up with birthday messages for Mary, a firefighter who is about to turn 50''} \\
    &
    \small b)
    & \small Cup problem
    & \small 
    Problem solving
    & \small \cite{p1225-yu.pdf,1214yu.pdf} %
    & \textit{
    \small ``How can you dry many cups quickly so that they don’t take up too much space [...]?''}\\
    &
    \small c)
    & \small Sketching
    & \small Artistic creativity
    & \small \cite{sheepmarket,p1393-yu.pdf} 
    & \small \textit{``Draw a sheep facing left''} or \textit{``Please design a chair for children''}\\
    \midrule
    \multirow{2}{*}{\small
        \rotatebox[origin=c]{90}{
            \makecell[c]{Creat.\\tests}
        }
    }
    &
    \small d)
    & \small Alternate Uses
    & \small Divergent thinking test
    & \small \cite{Gerber_AffectiveComputationalPriming.pdf,MUM2019} 
    & \small \textit{``[...] 
    think  of  as  many  unique  and unusual  uses  for  a  common  object''} \\
    &
    \small e)
    & \small Remote Associates
    & \small Convergent thinking test
    & \small \cite{huang2015.pdf,8a01bb36375a507fa01ad95e2eb83ebf2f65.pdf}
    & \small \textit{``find a word that [is] logically linked to the set of three words''}\\
\bottomrule%
\end{tabularx}%
\caption{Examples of creative tasks and creativity tests given to crowd workers on crowdsourcing platforms.}~\label{tab:exampleHITs}%
\end{table*}%
%
%
%
%
%
\section{Background}%
Our work is scoped within the intersection of creativity-oriented research in HCI and
1)
paid crowdsourcing platforms as an increasingly popular source of participants for research studies, and 
2)
creative work on these platforms.%
%
%
%
\subsection{Creativity-oriented Research in HCI}%
Historically, Guilford's presidential address to the American Psychological Association in 1950 \cite{GuilfordPresidentialAddress} launched the
``first wave'' of creativity research~\cite{p1235-frich.pdf}.
This era was dominated by the thought of creativity as a lone individual's private struggle.
In the tradition of research on intelligence,
psychometric tests were developed in the following decades 
to measure divergent and convergent thinking, as two 
determinants of creativity~\cite{GuilfordPresidentialAddress}.

Creativity is, however, a multi-faceted concept and hard to define and measure.
\rev{
Research on creativity roughly falls into two camps: 
H-creativity and P-creativity~\cite{boden}.
The former focuses on eminent creativity (or Big-C), that is, creative contributions of historical significance to society~\cite{Csikszentmihalyi}.
The latter is concerned with everyday creativity 
and novel insights at the individual level.
Besides Big-C, Kaufman and Beghetto's four C model distinguishes between
Pro-C (creativity at a professional level),
little-c (everyday creativity with focus on the resourcefulness of ordinary people), and
mini-c (i.e., the individual's learning process)~\cite{Beyond_Big_and_Little_The_Four_C_Model_of_Creativi.pdf}.}
Most definitions of creativity used in research are influenced by the Big-C and Pro-C perspectives.
Following this sociocultural view of creativity~\cite{Csikszentmihalyi}, the outcome of a creative activity must 
be
both original (novel, unusual, or unique) and effective (valuable, useful, or appropriate). 
This ``standard definition'' of creativity
(i.e., in this case divergent thinking)
has become a staple in the toolbox of many researchers~\cite{2012RuncoJaegerStandardDefinition.pdf}.
Other commonly used measures of divergent thinking include fluency (i.e., quantity  of  ideas),  flexibility  (i.e., number  of  different categories of ideas), originality (i.e., statistical infrequency of an idea), and practicality~\cite{p357-kerne.pdf}.
Some researchers (e.g., Amabile \cite{Amabile:1990}) argued that it does not matter how creativity is defined, because allowing participants to use their own definition may help them be more consistent.

In HCI, supporting creative work has been regarded as one of the field's grand challenges~\cite{Shneiderman2009}.
Current creativity-oriented research in HCI (the ``second wave''~\cite{p1235-frich.pdf}) is characterized by collaboration as a research theme in a paradigm shift towards studying creativity as an attribute of groups. 
This increased interest within HCI
has resulted in the development of a plethora of creativity support tools (CSTs) that aim to augment the creativity of groups and individuals~\cite{2019_chi-paper.pdf}.
%

\todo{As with any system,} quantifying the effectiveness of creativity support tools is important.
To this end, Shneiderman organized a workshop in 2005 on evaluating creativity support tools~\cite{shneidermanWorkshop}.
Since then, various metrics have been developed to measure a system's ability in providing creativity support~\cite{p127-carroll.pdf,a21-cherry.pdf,a14-kerne.pdf,ECQ}.
Some of these approaches focus on the experience of creativity.
For instance, the Creativity Support Index (CSI)~\cite{p127-carroll.pdf,a21-cherry.pdf} aims to measure how well a CST supports creativity and the Experience of
Creativity Questionnaire (ECQ)~\cite{ECQ} measures experiential and existential aspects of artistic creativity.
\rev{Both the CSI and ECQ aim to give an insight into how creative work is experienced \rev{(in the sense of little-c and mini-c creativity)}.}





\subsection{Paid Crowdsourcing Platforms as Participant Pools}%
%
Crowdsourcing platforms have become popular sources of participants in research~-- in and outside of the field of HCI.
For instance, studies in behavioral research~\cite{Mason-Suri2012_Article_ConductingBehavioralResearchOn.pdf} and cognitive science~\cite{PIIS1364661317301316.pdf} have experienced a significant uptake in the use of 
MTurk in recent years.
Workers on MTurk have become one of the most thoroughly studied sets of human subjects~\cite{annurev-clinpsy-021815-093623.pdf}.
For this reason, scientists have at their disposal a strong understanding of the demographics and availability of 
workers on this crowdsourcing platform (e.g., \cite{p135-difallah.pdf,p16-ipeirotis.pdf,ce5288bb36bf7b636f709c00e00e22a02d07.pdf,Paolacci2014}).
Other platforms transparently report demographic data themselves (e.g., 
\cite{ProlificDemographics}).

The extensive use and, some will argue, over-reliance on CS platforms as a convenient mechanism for data collection has sparked criticism. For instance, Anderson et~al. refer to this burgeoning phenomenon as the ``MTurkification'' of research~\cite{anderson2018.pdf}.
Data collection on crowdsourcing platforms may be affected by what Stewart et~al. refer to as an ``emerging tragedy of the commons''~\cite{PIIS1364661317301316.pdf}.
A large portion of the tasks are completed by a relatively small pool of active professional workers who have been exposed to many types of different tasks
\cite{chandler2013.pdf,p135-difallah.pdf,PIIS1364661317301316.pdf,jdm14725.pdf}.
A high number of crowdsourcing tasks may be carried out by 
the most active ``professional Turkers''~\cite{Caution-MTurk-Workers-Ahead-Fines-Doubled.pdf} or ''super Turkers''~\cite{bohannon2011.pdf,chandler2013.pdf}.
There may be a substantial overlap between the populations on MTurk accessed by different laboratories~\cite{jdm14725.pdf}.
Naturally, the overlap is not limited to human-subject research in laboratories but can occur within a domain, such as creative work.%
%
%
%
\subsection{Creative Work on Crowdsourcing Platforms}%
%
Following the view of information-based ideation, 
creative work can be defined as ``open-ended tasks and activities in which users develop new ideas''~\cite{a14-kerne.pdf}.
But ideation (i.e., generation of ideas) is only one type of creative work.
Table~\ref{tab:exampleHITs} lists selected examples of other types of creative work that were given as tasks to online workers on crowdsourcing platforms.

Two common types of such creative work are eliciting creative input for a given purpose in \textit{creative tasks} (see Table \ref{tab:exampleHITs} a--c) and studying creativity itself in \textit{creativity tests} (Table \ref{tab:exampleHITs} d and e).
Psychometric creativity tests measure a subject's creative potential and can be classified into tests of divergent thinking (e.g., the Alternate Uses test~\cite{guilford1978alternate}) and convergent thinking (e.g., the Remote Associates test~\cite{RAT}).
Creative tasks, on the other hand, ask crowd workers to contribute to ideation in one way or another by providing creative ideas to a crowdsourcing campaign. 

\rev{From the requester's perspective,
creativity tests are focused on measuring people's creativity,
whereas creative tasks elicit people's creative output.
In research, creative tasks are also often used to determine people's creativity. 
For example, Siangliulue et~al. asked workers to generate birthday greetings~\cite{IdeationAtScale-cscw2015.pdf} and Yu et~al. asked workers to generate creative solutions to a given problem~\cite{1214yu.pdf}. In both studies, 
expert judges evaluated the creativity of the generated ideas, resulting in a measure of people's creativity.
Expert judgment is a technique that has been applied in a broad range of studies~\cite{The_Cambridge_Handbook_of_Creativity.pdf}.
Creativity tests also may involve judgment along a number of criteria (idea fluency, flexibility, etc.). The same criteria can be found in many studies evaluating creative tasks.}
Creativity tests can further be used for measuring other constructs.
For instance, Lu et~al. used the Remote Associates test for measuring unethical behavior of participants~\cite{8a01bb36375a507fa01ad95e2eb83ebf2f65.pdf}.
From the crowd worker's perspective, however, the experience of creative tasks and creativity tests can be remarkably similar (see Table~\ref{tab:exampleHITs}).
\rev{Workers may not even be clear if there is a difference at all, as oftentimes requesters are 
not transparent about the aims of their study (e.g., to prevent bias in study participants).}
\rev{Taking the perspective of the worker, we view creativity tests and creative tasks as two subsets of creative work on crowdsourcing platforms in this paper.}


Concerning studies that involve eliciting creative work, some researchers have raised doubt about the applicability and effectiveness of paid crowdsourcing platforms, such as Amazon Mechanical Turk~\cite{dontcheva.pdf,p22-kittur.pdf,Gerber_AffectiveComputationalPriming.pdf}.
For instance, Gerber et~al. anecdotally raised concern that MTurk ``may not be the best platform for creativity studies''~\cite{Gerber_AffectiveComputationalPriming.pdf}.
These platforms were originally created for highly parallelizable tasks such as image labeling or text annotation. MTurk, in particular, ``was not designed with creative tasks in mind''~\cite{dontcheva.pdf}.
Microtask crowdsourcing is especially suitable for short tasks that incur a low cognitive load and are objectively verifiable~\cite{p22-kittur.pdf}.
Creative tasks, on the other hand, are subjective and there is no right answer.
In the absence of a verifiable ground truth in creative tasks, quality control becomes a challenge.
Open-ended, subjective tasks may be vulnerable to \todo{exploitation} by 
workers~\cite{p1631-gadiraju.pdf}.
Further, creativity itself is a multi-faceted concept that is hard to define and measure precisely and
the HCI literature lacks a unified definition for creativity~\cite{p1235-frich.pdf}. 

Other identified issues exist that could affect the validity of creativity studies on crowdsourcing platforms.
For instance, prior research has found evidence of worker nonna\"{i}vet\'{e}~\cite{chandler2013.pdf}. A worker’s prior exposure to commonly used manipulations and measures may negatively impact the {validity} 
of the results obtained~\cite{chandler2013.pdf,chandler2015.pdf,194.pdf}.
Specific to creativity studies, foreknowledge~-- either as a result of prior exposure to creativity studies on the crowdsourcing platform or having learned about creativity tests in formal education -- may lead the worker to not reflect upon the work and recall answers from memory instead of using their creative ability.
In addition, nothing stops workers on crowdsourcing platforms from turning to the Web to search for answers. 

\todo{
As CS platforms conveniently allow recruiting participants for creativity studies and eliciting creative work,
CS platforms have become excellent means for 
creativity-oriented research and 
potential sources for supporting creativity in \textit{crowd-powered creativity support tools}~\cite{DC2S2:2019}.}
In our study, we provide an in-depth worker-focused look into the space of using crowdsourcing platforms in creativity-oriented research.
We look into worker nonna\"{i}vet\'{e} in regard to the most commonly used creativity tests and touch on a number of issues previously reported in the literature, such as underpayment of workers~\cite{Irani:2013,responsible-research} 
and worker motivation~\cite{More_than_fun_and_money_Worker_Motivation_in_Crowd.pdf,Malone}. 
As a result of our studies, we profile workers specifically concerning creative work, and provide much-needed qualitative insights into how the workers themselves perceive creative work on paid crowdsourcing platforms.

\section{Study Design}%
%
We published a worker-focused online questionnaire as a task on two different crowdsourcing platforms.%
%
\subsection{Choice of Crowdsourcing Platforms}%
%
Our work is scoped to two popular crowdsourcing platforms on which workers self-select to work on paid tasks~\cite{howe2006}.
Not included in the scope of the study are online platforms that were specifically created for outsourcing creative tasks, such as
Upwork,
Fiverr,
DesignCrowd,
99Designs,
and Innocentive.

We selected Amazon Mechanical Turk (www.mturk.com) and Prolific (www.prolific.co) for our study.
\textit{MTurk} is likely the most popular general-purpose crowdsourcing platform
for requesters to distribute ``Human Intelligence Tasks'' (HITs) to an anonymous crowd of workers. HITs are typically small units of work 
that are too difficult for machines to solve, but can be completed quickly by humans. Examples of HITs are image transcription, sentiment analysis, and gathering information in online searches.
\textit{Prolific} is a crowdsourcing platform targeted towards academic studies~\cite{beyond-the-turk-an-empirical-comparison-of-alternative-platforms.pdf}. Prolific was primarily created for behavioral, user, and market research. Studies posted on Prolific are often online surveys eliciting personal viewpoints on a topic, but may also, for instance, include complex online experiments and mobile application research.

The two platforms are different in their demographics and the type of work offered. Both platforms can provide excellent results for their respective use cases.
Our aim in this paper is not to compare the two platforms, but to provide complementary insights into what creative work is like on two of the most popular crowdsourcing platforms used in HCI research.

\subsection{Questionnaire and Procedure}

The questionnaire consisted of 34~items,
including an instructional manipulation check 
(IMC) 
and an attempt to identify inattentive or non-serious participants.
The dominance of collaborative creativity in the current HCI literature prompted us to include inquiries about collaboration in our survey.
The full questionnaire is available in the Auxiliary Material of this paper.

The questionnaire was published
in six batches 
(one batch per day, Monday to Saturday) and during different times of day (in EDT: 1~am, 5~am, 9~am, 1~pm, 5~pm, and 9~pm). Participants were rewarded with US~\$1 on MTurk and 
UK \pounds1 on Prolific.

After providing their consent, participants were eased into the study with questions about their personal experience and working preferences on the respective platform. Next, we asked the workers to define creativity in their own words and to provide us with their view of creative work on the respective platform. 
After they provided their open-ended thoughts on creativity and creative work,
the participants were anchored to the following descriptions for the remainder of the study:
\begin{quote}%
    \textit{Throughout this questionnaire, by "CREATIVITY STUDIES" we mean surveys and tasks on <platform> that
    \begin{inparaenum}[1)]
        \item test your own creativity (\textbf{``creativity tests''}), or
        \item ask you to be creative (\textbf{``creative tasks''}), e.g. generate ideas.
    \end{inparaenum}
    An example of a creativity test might be where you're given an item and asked to come up with creative uses for it. Or you are given a few words and asked to come up with a related word.\\
    An example of a task which asks you to be creative would be thinking of names for a new app, or writing an article for a blog, or designing a logo for a new car.}%
\end{quote}%
The subsequent questions either referred to creativity studies as a whole, or inquired in more depth about creativity tests and creative tasks individually.%
%
%
%
%
%
%
\subsection{Participant Screening}%
In total, we recruited 323~participants (170~from MTurk and 153~from Prolific).
The discrepancy in numbers was caused by workers dropping out or timing out on the task. We republished the survey as necessary 
to make up for this attrition.

The qualification criteria for the two crowdsourcing
campaigns were as follows.~On MTurk, we required workers to have completed at least 1000~HITs and have a HIT approval rate of 98\% or higher. Similar qualification criteria are typically being used in academic studies to improve the chance of receiving high quality responses~\cite{peer2013.pdf}.
On Prolific, participants were required to have an approval rate of at least 98\% and to be fluent in English.
On MTurk, fluency in English is expected by default, requiring no extra action from the requester~\cite{Q14-1007.pdf}.
Participants were further asked whether they had \textit{``participated in studies about creativity (e.g., creativity tests), or <HITs/studies> that demand creative thinking, on <MTurk/Prolific> in the past.''} Participants who answered ``yes'' or ``maybe'' to this question were directed to the rest of the questionnaire.
Participants who responded ``no'' to this question ($N=82$; 25.4\%) were redirected to an unrelated online experiment (not in the scope of this paper) and not considered in the study.%

A further 26~participants were removed from analysis because they failed the qualification tests or tried to game the survey.
Of these participants, five MTurk workers had each taken the survey twice.
Nine MTurk workers failed to answer the attention check (\textit{``What planet do we live on''}) correctly. None of the participants on Prolific failed this question.
Fifteen of the participants (14 from MTurk, 1 from Prolific) failed the instructional manipulation check.
Some workers failed both the IMC and the attention check.
%
The final set of 215~participants consisted of 102~MTurk workers and 113~members of Prolific.%
%
%
%
%
%
\subsection{Qualitative Analysis Methodology}%
%
We coded the responses to the open-ended questionnaire items following Clarke and Braun's recommendations for thematic analysis~\cite{braun2006.pdf}, with modifications as follows.
The first author, knowledgeable in research on creativity, first 
open-coded 30\% of the responses.
The emerging initial set of codes
was then shared and discussed with two other coders.
The code set was refined and extended by the group. Each code was annotated with several representative examples.
The three researchers then individually applied the codes to the data
and used the constant comparison method~\cite{Glaser_1967.pdf} to iteratively improve the set of codes while coding. All coders were blind to the worker's crowdsourcing platform.
The group met two times to share updates to the codes and differences were resolved in discussions. Two rounds of coding were conducted.
Inter-rater reliability was not calculated, as disagreement among the researchers was resolved through critical and detailed discussions and the process involved multiple rounds of coding~\cite{McDonald_Reliability_CSCW19.pdf}.%
%
%
%
%
%
\section{Data and Findings}%
On average, participation took approximately 15~minutes 
on MTurk and 9.5~minutes 
on Prolific. Noticing our estimated task completion time being too low, and in accord with the guidelines of fair crowd work~\cite{responsible-research}, we compensated a subset of the participants with bonuses so that everyone earned a minimum of US~\$7.50 per hour.%
%
%
\subsection{Worker Demographics}%
%
\textit{Age and gender identity:}
The workers in our sample had similar age ranges.
MTurk workers were aged 21--71 ($M = 33.3$~years, $SD = 8.9$~years), while 
workers from Prolific were aged 19--72 ($M = 37.1$~years, $SD = 11.6$~years). 
The participants included 128 men (66 on MTurk, 62 on Prolific) and 87 women (36 on MTurk, 51 on Prolific). None of the participants identified with a non-binary/third gender, and none of them opted to enter their own gender identity or to keep it 
private.

\textit{Location:}
The majority of the MTurk workers came from the United States (66\%) and India (32\%). A negligible amount of MTurk workers were located in other countries (one in the UK, one in Italy).
The Prolific sample was more geographically diverse, with peak participation of 40.7\% in the United Kingdom, followed by the United States (31\%), Portugal (5.3\%), and Poland (4.4\%), among other countries.

\textit{Education:}
About half of the MTurk workers (52 participants) 
had a Bachelor degree, compared to about one third (40 participants, 35.4\%) on Prolific. Equal portions of participants on both platforms (14 participants, 13.7\%, on MTurk versus 15 participants, 13.3\%, on Prolific) held a Master's degree. The Prolific sample contained about twice as many participants with a college degree (17 participants, 15.0\%,  vs. eight participants, 7.8\%, on MTurk) and more participants with a high school degree (25 participants, 22.1\%, vs. 18 participants, 17.6\%, on MTurk). One participant from Prolific had a doctoral degree.

\begin{figure}[htb]
\centering%
\includegraphics[width=\columnwidth]{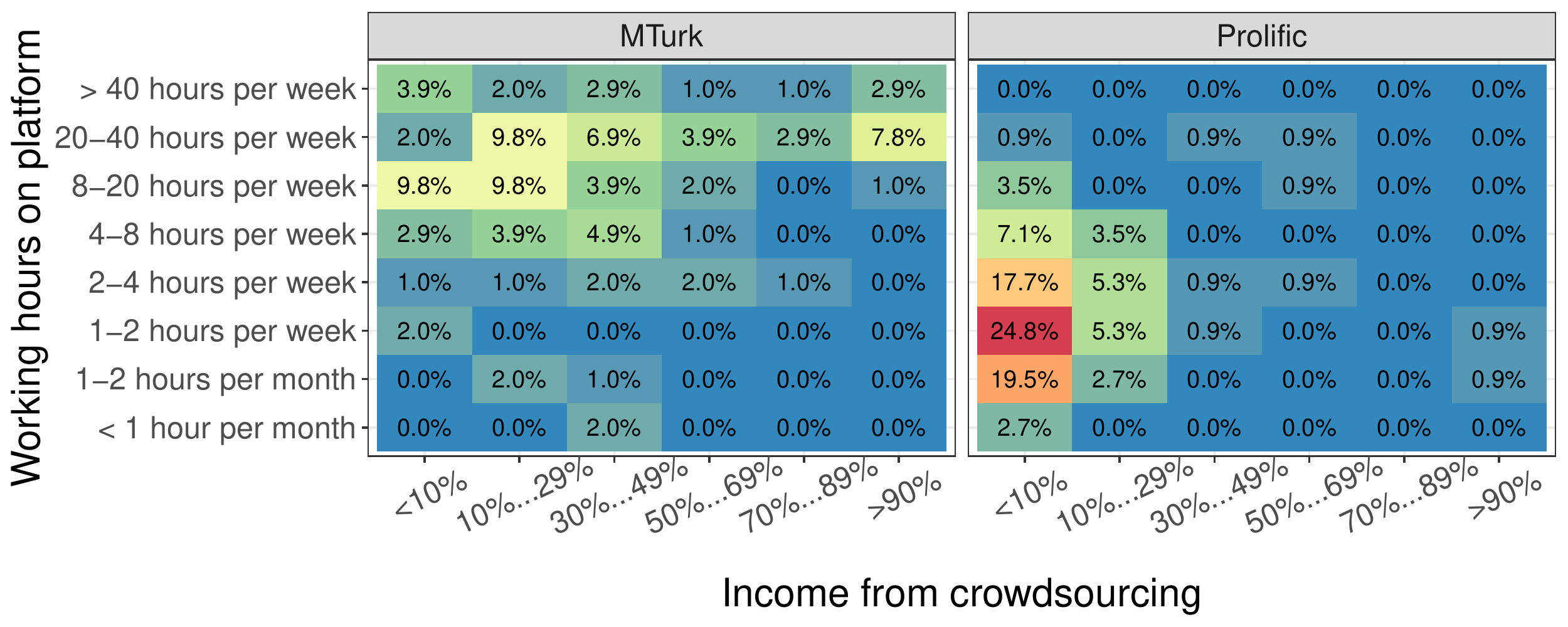}%
\caption{Worker responses to two multiple choice questions: working hours on the respective crowdsourcing platform versus percentage of the worker's income from crowdsourcing (rounded to one decimal).}%
\label{fig:scatter}%
\end{figure}

\textit{Employment status and weekly work hours:}
%
The participants from Prolific were a much more casual workforce than the MTurk workers (see the weekly work hours and the income in Figure~\ref{fig:scatter}).
About 70\% of the MTurk workers (72~participants) said they work full-time (40+ hours) on the platform compared to around half of the Prolific workers (58~participants). Fourteen MTurk and 16~Prolific participants worked part-time. About twice as many Prolific participants were self-employed or worked from home (17~participants, 15\%) compared to the MTurk workers (7~participants, 6.9\%).

\textit{Income from crowdsourcing:}
A Chi-square test confirmed that the workers from the two CS platforms had significantly different income distributions, $\chi^2(5)=71.9$, $p<0.01$.
Most notably, about three quarters of the participants on Prolific ($N=86$) make less than 10\% of their income from their activities on the crowdsourcing platform. On MTurk, only about one in five workers responded this way ($N=22$; 21.6\%).%

\textit{Work experience:}
%
Our sample captured a wide range of workers, from part-time workers with a short work history to seasoned full-time workers with a long-standing affiliation with the platform.
MTurk workers had, on average, a significantly longer affiliation with the crowdsourcing platform ($M = 29.4$ months, $Mdn=24$ months) compared to the members of Prolific ($M = 16.9$, months, $Mdn=12$ months, $t(137.1) = 4.58$, $p<0.01$).
However, with a minimum of 107 completed approved studies ($Max=1210$, $M=259.3$, $Mdn=182$) and a low number of rejections ($M=1.5$, $SD=1.8$), 
the Prolific participants in our sample were not novices.%
%
%
%
\subsection{Motivation of the Crowd Workers}%
Without mentioning creative work, we first inquired about the workers' motivation for completing tasks on the respective crowdsourcing platform. The responses provide further evidence of the general trend in our data that members of Prolific have a more casual approach to work.
We note that MTurk workers work much longer hours on the platform than participants from Prolific (see Figure~\ref{fig:scatter}).
Approximately 70\% of the participants from Prolific find working on Prolific a \textit{``fruitful way to spend free time and get some cash,''}
compared to only one third (34\%) of the MTurk workers.
Similarly, 41\% of the MTurk workers earned their primary income from MTurk, while only 14.7\% of Prolific workers claimed to do so.
The majority (84\%) of the workers on Prolific earned ``secondary'' income from their platform, compared to 54\% of the MTurk workers.
About a quarter (23.5\%) of the Prolific members worked on the platform \textit{``to kill time,''} while only approximately every tenth MTurk worker responded this way.
Participants on Prolific also thought the tasks were more entertaining to complete.
Almost two thirds of the Prolific participants (62.7\%) thought the tasks were \textit{``fun,''} compared to only one quarter (24.5\%) of the participants on MTurk.
%
%
In summary, we note there are significant differences in the demographics and work preferences on MTurk and Prolific. Next, we turn to investigating creative work on the two platforms.%
%
%
%
%
%
%
\subsection{Creative Work Through the Lens of a Worker}%
A subset of the participants ($N=152$) proceeded to provide open-ended reflections on creative work on the respective CS platform.
Participants primarily expressed {opinions on the amount of current tasks} ($N=42$; 27.6\%) and mentioned details about the {enjoyment of creative work} ($N=24$; 15.8\%). As for amount, the most common recurring element was the low availability of creative work on the platform: \textit{``I think creative work is very uncommon on MTurk, as most of the work is very technical''} (P184, MTurk),
or \textit{``Prolific is almost entirely academic studies [...] so I don't think there's really creative work here in the first place''} (P44, Prolific).
Enjoyment of creative work, while mentioned often, also may come with certain reservations: \textit{``I think creative work is fun, but I only do creative or subjective work for requesters who are known to be fair requesters -- I have suffered unfair rejections just because a requester didn't like my answers''} (P145, MTurk).
Inspiration was mentioned as one of the benefits of creative work ($N=18$; 11.9\%). Creative work was seen as \textit{``a great way express your feelings''} (P68, Prolific) and \textit{``opinions''} (P95, Prolific). Creative work \textit{``always makes you think''} (P109, Prolific) and \textit{``keeps the brain and mind active''} (P30, Prolific).
Analyzing the responses, it also became clear that the participants did not at this point think about creativity tests: they only considered
tasks such as content creation or tasks that allow for a degree of freedom in the work itself.%
\begin{figure*}
\centering%
\begin{subfigure}{0.9\textwidth}
\includegraphics[trim={0.9cm 0 0 0},clip, width=\textwidth]{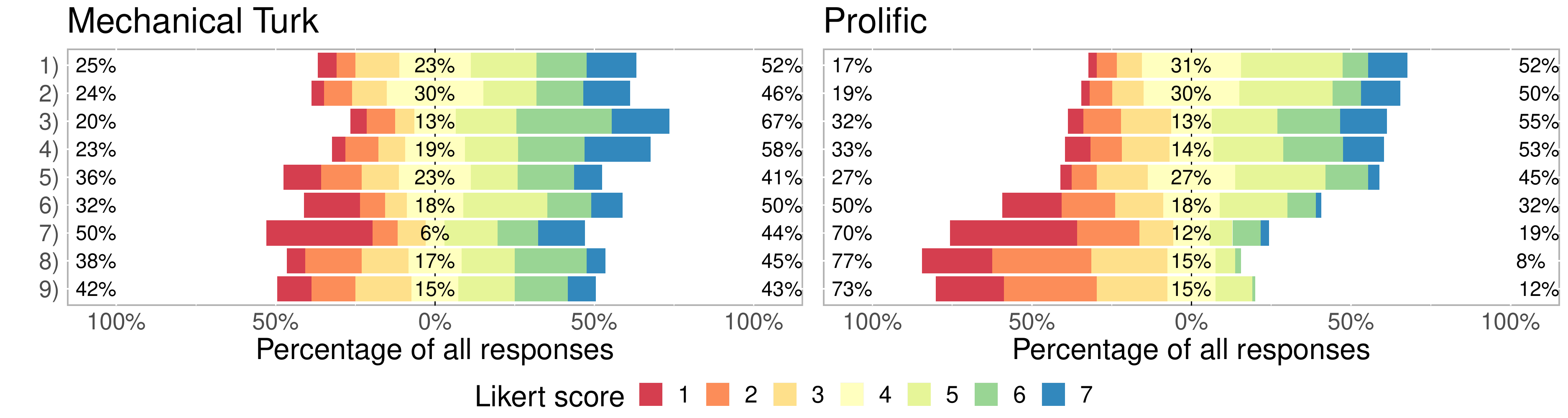}%
\end{subfigure}\\%
\begin{subfigure}{\textwidth}%
\centering%
\vspace{2mm}%
\begin{tabularx}{0.92\textwidth}{>{\hsize=0.01cm}Xl}
\scriptsize 1) & \scriptsize How many creative tasks should there be on <PLATFORM>? (1: Much less, 7: Much more) \\%
\scriptsize 2) & \scriptsize How many creativity tests should there be on <PLATFORM>? (1: Much less, 7: Much more) \\%
\scriptsize 3) & \scriptsize Overall, the creative tasks on <PLATFORM> you participated in were... (1: Not at all interesting, 7: Extremely interesting) \\%
\scriptsize 4) & \scriptsize Overall, the creativity tests on <PLATFORM> you participated in were... (1: Not at all interesting, 7: Extremely interesting) \\%
\scriptsize 5) & \scriptsize What type of tasks do you generally prefer on <PLATFORM>? (1: Simple and easy tasks, 7: More complex tasks that make me think) \\%
\scriptsize 6) & \scriptsize Looking at your past participation in creativity studies on <PLATFORM>, how much do you think you learned about yourself? (1: Nothing at all, 7: Extremely much) \\%
\scriptsize 7) & \scriptsize “I first learned about creativity tests on <PLATFORM>.” 
(1: Strongly Disagree, 7: Strongly Agree) \\%
\scriptsize 8) & \scriptsize How often have you seen creative tasks being offered on <PLATFORM>? (1: Not at all often, 7: Extremely often) \\%
\scriptsize 9) & \scriptsize How often have you seen creativity tests being offered on <PLATFORM>? (1: Not at all often, 7: Extremely often) \\%
\end{tabularx}%
\end{subfigure}%
\caption{Worker responses to the 7-point Likert items on various perceptions of creative work. The percentages on the left, middle and right indicate disagreement (1--3), neutrality (4) and agreement (5--7), respectively.
For example, 77\% of the Prolific workers indicated that they did not see creative tasks often on their crowdsourcing platform, while 8\% encountered them more frequently.}%
\label{fig:likertgraphs}%
\end{figure*}%
%
%
\subsection{Past Encounters of Creative Work}%
%
One of our initial assumptions was that workers
on {paid crowdsourcing platforms} are getting used to seeing the same type of creative work over and over.
To this end, we asked the participants who
had seen creative work on the platform to describe one such instance.
Of the 167~workers who articulated such an encounter,
one third ($N=56$; 33.5\%) articulated one of the standard creativity tests for measuring divergent and convergent thinking:
the Alternate Uses test (36~mentions; 21.6\%) and the Remote Associates test (7 mentions; 4.2\%). 
This is a remarkably high proportion of workers~-- and most likely an underestimation since other workers might have taken standard creativity tests as well but did not recall and articulate one here, as they were only requested to describe the first instance of creative work that came to their mind.
Over two thirds ($N=25$; 69.4\%) of the participants who had taken the Alternate Uses test mentioned they had been given a brick as an object.
Besides creativity tests, the second most popular type of creative task was producing content (text/graphics) ($N=44$; 26.3\%), followed by ideation tasks ($N=29$; 17.4\%), such as \textit{``Coming up with ideas for marketing a product''}~(P15, Prolific).

We further inquired how workers had been first exposed to creativity tests.
Almost one third ($N=51$; 32.7\%) of the participants who provided an answer to this optional question ($N=156$) had learned about and taken such tests
in their formal education at high school, college or university. Others had first encountered creativity tests on the crowdsourcing platform ($N=21$; 13.5\%), other platforms/websites ($N=16$; 10.3\%), or just elsewhere ($N=41$; 26.3\%). Less than one in five respondents 
($N=27$; 17.3\%) reported they had never heard about creativity tests before.
As a particularly interesting anecdote, one participant (P212, MTurk) mentioned: \textit{``Stop repeating the task! I have answered the `come up with creative uses for a brick' question about fifty times. Pick something else!!!''}.
Another participant (P197, MTurk) noted in the same vein: \textit{``Some of the tasks look like they are classic textbook tests. People who do lots and lots of surveys and work on MTurk a lot will end up seeing these over and over, thus maybe invalidating any value the platform offers to requesters.''}%
%
%
%
\subsection{Quantitative Insights into the Creative Work Preferences}%
The comparison of attitudes and work preferences on 7-point Likert scales (depicted in Figure~\ref{fig:likertgraphs})
highlights the differences between the workers on the two platforms in regard to creative work.
Within Prolific, workers indicated they have not seen many creative tasks or creativity tests ($M=2.6$ and $M=2.7$, respectively), but they would like to see more in the future (tasks $M=4.5$, tests $M=4.6$), both confirmed by t-tests, $p<0.01$. Between the two platforms, Prolific workers have learned less than MTurk workers about themselves during past creative work ($M=3.4$ and $M=4.1$, respectively, $t(202.7)=2.7$, $p<0.01$). Further, MTurk workers had seen more creativity tests and creative tasks on the platform in the past than respondents from Prolific (creativity tests: $M=4.0$ vs. $M=2.6$, creative tasks: $M=4.1$ vs. $M=2.7$), t-tests, $p<0.01$. Workers from both platforms express equal preferences when it comes to desired complexity of work ($M=4.0$, MTurk and $M=4.2$, Prolific, where 1 meant
low complexity and 7 high complexity).
Finally, MTurk workers stated they had their first encounter with creative work on the crowdsourcing platform more often than Prolific workers ($M=3.6$ vs. $M=2.6$), $t(191.8)=3.4$, $p<0.01$.%
%
%
%
%
\subsection{Qualitative Insights into the Creative Work Preferences}%
Approximately half of the participants on both platforms (see Figure~\ref{fig:likertgraphs}) wished to see more creative work offered (with only 17\%--25\% wishing to see less).
%
Out of the workers who wanted less creative work, a majority (70\%) of participants did not, or could not, articulate a clear reason for their preference. Not surprisingly, among the workers who elaborated on their preference, the number one reason for wanting less creative work was monetary rewards, followed by 
creative work being \textit{``too complex''} and \textit{``difficult''} to complete.
The most common reasons for wanting more creative work to be offered on the platform were to introduce variety to the available work ($N=36$; 16.7\%) and, second, increase the enjoyment of work in general ($N=26$; 12.1\%). Or, as a member of Prolific (P16) articulated it: \textit{``It’s something I would enjoy rather than endure!''} Others mentioned positive aspects, such as learning 
during the creative processes and challenging themselves with work that requires \textit{``to get your brain stimulated''}~(P108, Prolific) rather than just completing \textit{``monotonous tasks.''}


We further inquired about the preference for working alone versus working collaboratively. In our sample the verdict was clear: 192~participants (89.3\%) preferred solo work and only 23 (10.7\%) preferred collaborative work.
We isolated various reasons for this preference by asking the participants to elaborate on their answers. First, cooperation issues ($N=37$, 17.2\%) were often brought up, with productivity and efficiency ($N=28$, 13\%) being the second most popular reason for workers to prefer solo work. Enforced cooperation causes issues both in time and rewards, but a few participants made remarks about the perceived skills gap between workers,
for instance \textit{``I prefer working alone since not everyone is on the same skill level as I am.
I don't want to be hindered or hinder anybody.''} (P187, MTurk).
Yet, there is a small, but clear, group of workers who enjoy collaborative creative work. For instance, P74 (Prolific) noted: \textit{``The tasks that include others tend to be more exciting due to the anticipation of seeing how they will respond to each task.''} Others informally noted that they are simply a `people person' and enjoy the company of other people -- regardless of the medium.%

Finally, we asked about what type of creative tasks the workers would want to see offered on the platform and why.
From the onset of analysis, we noticed a significant carryover effect from the previous item, as 50~participants (23.3\%) emphasized they wish to see creative work that can be completed alone and 10~participants (4.7\%) said they want to collaborate. Thirty-four participants (15.8\%) were largely indifferent or said \textit{``anything goes.''}
A particularly pragmatic approach to work was articulated as \textit{``I view this platform as a good research tool so I don't see my preferences as relevant. The fundamental question is what are the research needs''} (P53, Prolific). 
Some of the specific types of creative work mentioned were different types of ideation ($N=9$; 4.2\%), content production tasks ($N=8$; 3.7\%), and creativity tests that 13~participants (6\%) wanted to see more of.%
%
%
%
%
%
\subsection{Learning through Creative Work}%
We found evidence that for many of the participants, learning is a reason for doing creative work.
The proportion of participants reporting on some aspect of learning taking place during creative work on the crowdsourcing platform is high.
Of the 143 workers who elaborated on what they learned (if anything),
only six (4.2\%) reported not learning anything at all.

About thirty percent of the participants ($N=43$) mentioned that participation in creativity studies has led to discovering insights about their own personality.
One fifth of the workers ($N=28$; 19.6\%) reported either developing their creative skills or awakening to the fact that they already are creative: \textit{``I learned that I have more problem solving skills than I give myself credit for''}~(P107, Prolific).
Others mentioned having learned subject-specific skills ($N=17$; 11.9\%). Increasing one's productivity ($N=7$; 4.9\%) and concentration were mentioned ($N=6$; 4.2\%): \textit{``I learned that I can be pretty creative even while under time pressure''} (P201, MTurk) and \textit{``I learnt that I can concentrate for long periods of time''} (P106, Prolific).
Working on the platforms seems to encourage workers to engage and test their creativity in other areas, as exemplified by P132 (MTurk) who \textit{``learned how to write scripts to reduce time required to work a Hit.''}%
%
%
%
\subsection{Specific Worker Concerns}%
Finally, we asked all participants to 
provide their view on currently existing or potential pain points and suggestions for requesters of creative work on the respective CS platform.
Most of the workers ($N=142$; 66\%) did not have anything in mind, replying simply with a form of `no.'
Among the 
remaining responses, a variety of different perspectives emerged.%

Unsurprisingly, \todo{24}~respondents \todo{(11.2\%)} explicitly mentioned money and rewards as an issue in creativity studies.
Most often, the workers were concerned about creative work underpaying participants. 
P133 (MTurk) noted that \textit{``creative tasks do not pay enough and exclude too many people on too many tasks. There are better platforms for these sorts of things.''} Time consumption and low reward are closely related factors to consider for workers who may think that creative tasks are \textit{``just not worth the time or effort for the money they offer''} (P170, MTurk).
Clearly, there is a trade-off between individual fulfillment by using one's creativity and working income. As P206, who had been working for 3 years on MTurk and was currently working full-time on the platform, put it: \textit{``I want to have more creative hits but the money comes from the boring hits.''}

Fifteen participants \todo{(7\%)} voiced concerns about creative work taking too much time to complete.
Among these participants, creative work was perceived as stressful: \textit{``Sometimes you have only one or two minutes to find new ideas which is kind of stressful''} (P31, Prolific).
Participants noted that creative tasks \textit{``are usually too long and complicated''} (P98, Prolific), \textit{``really drawn out''} (P100, Prolific),
and \textit{``take more time than estimate, and therefore are not worth the time for the monetary return''} (P13, Prolific).
%
For some workers, the challenge 
is more in trying to \textit{``avoid boring tasks''} (P102, Prolific). Finding alternate uses for a brick, for instance, was seen as \textit{``stupid''} (P203, MTurk).
One participant (P146, MTurk) proclaimed: \textit{``I've never seen [a creative task] that was worth my time. They are all uninteresting by their nature.''}

Other workers complained about creative tasks being \textit{``difficult''} \todo{($N=5$; 2.3\%)} to complete: \textit{``sometimes the restrictions are too much to really get creative, I end up just giving what they are looking for: carbon copy results''} (P158, MTurk).
Open-ended tasks may, however, also be an opportunity:
\textit{``Creative tasks don't have a right or wrong answer and so as long as you follow the guidelines then your work is accepted. I like that because sometimes I wonder if my ideas are stupid but really I don't have to worry about that''}
(P105, Prolific).
P145 (MTurk) suggested to requesters:
\textit{``You should be VERY lenient about creative tasks. You should not reject work for subjective assignments unless it is clear and unequivocal that the worker was scamming the requester.''}

Various other 
viewpoints \todo{($N=30$; 14\%)} emerged.
Among these workers,
concerns about creative work being exploitative and about the ownership of ideas were raised
\todo{($N=12$ out of 30, 5.6\% of the 142 respondents)}: \textit{
``creative tasks seem more appropriate for sites/jobs that pay their authors/artists/creative workers an appropriate wage [...]
It feels exploitative to use a survey site under the 'disguise' of research''} (P32, Prolific).
Or as P178 (MTurk) put it: \textit{``I feel that my creativity is my own and if I wish to use I will.
I don't want to give my ideas to a person or company I don't know.''}
Lack of transparency on the aims of online studies and how the results of such studies will be used was mentioned as a concern by seven \todo{(3.3\%)} of the workers.%
%
%
%
%
\section{Discussion}%

Crowdsourcing platforms have emerged as vast potential sources
\rev{for participants and original thought.}
Most likely we have not yet unlocked the full potential of this combination, and there is much to discover about the feasibility of crowdsourced creativity for both scientific and industrial purposes.

But why does the workers' perspective matter to begin with? One can argue the workers on the crowdsourcing platforms have a choice: they do not need to opt for creative work. An equally fair argument can be formed by considering the worker--requester relationship. For instance, steps have been taken towards reducing the invisibility of workers in pursuit of ethical development of crowd work~\cite{WebSci-CrowdsourcingEcoSystem.pdf,Irani:2013}. Similarly, we identified evidence in our study of some workers wishing to see the final results of the conducted experiments, speaking toward some relation beyond just purchasing units of labor.

Distinct calls in HCI have been made for us, the designers and creators, to question our own role and actively develop toward an improved society for all~\cite{Bardzell:2010:FHT:1753326.1753521}. This includes the workers on platforms who we now commonly engage as participants in online studies or for harvesting data. Indeed, we can regard crowd work as inherently participatory: the labor obtained from crowdsourcing platforms is not purchased as cycles of human-computational labor produced by anonymous ``humans-as-a-service''~\cite{Irani:2013}, but it originates from stakeholders, even if paid ones, in a co-creation process. Acknowledging the needs and wishes of the stakeholders allows us to transition to a more humane relationship with the workers, who come as diverse as humans typically do. In our study, we could 
observe a variety of emerging worker archetypes.%
%
%
%
\subsection{Crowd Worker Archetypes}%
Based on the thematic analysis of the data and clustering by affinity, we formed an impression of the workers' behavior, attitudes, and preferences for creative work.
In the following, we discuss the worker archetypes that best describe our shared understanding of the different types of workers emerging in our study. We supplement the descriptions with 
selected statements from the participants themselves. Naturally, these archetypes are formed based on the authors' subjective understanding of the workers in the specific sample of the study, and we do not claim this to be an exhaustive, all-encompassing list.%
%
%
\subsubsection{The Professional Crowd Worker}
The most evident type we encountered is the worker who is ``in it for the money.'' 
The professional crowd worker completes tasks full-time and for long hours in pursuit of maximum productivity and income. The professional worker is an attentive worker~\cite{Hauser-Schwarz2016_Article_AttentiveTurkersMTurkParticipa.pdf} and knows about attention checks and IMCs.
This worker prefers working alone, as collaboration~-- and especially so in creative work~-- entails uncertainty, such as unpredictable technical problems, cooperation issues and precarious rewards that are difficult to estimate. This worker is not interested in experimenting and trying new things~-- not even improving oneself beyond productivity alone. Finally, the professional crowd worker has been exposed to common creativity tests and creative work in assignments on the crowdsourcing platform.

In our study, participants for the most part wanted to work alone and highlighted the need for work with high rewards. Or, as a participant from MTurk put it, when asked whether there should be more creative tasks available on the platform: \textit{``As long as the pay is reasonable for the work time, it does not really matter, we as workers just want to be paid reasonably for the time we put in''} (P200, MTurk).%
%
%
\subsubsection{The Casual Worker}%
The casual worker works for additional earnings and pocket-money. 
This type of worker will do repetitive work, if necessary, but prefers tasks that allow for imagination and creativity to flourish.
This worker type is not limited to students and unemployed or self-employed workers, but may be more prevalent in these groups that do not have a steady full-time income.
Of the different worker types, casual workers are the most open towards collaborative creative tasks and experiments, as the outcomes of their crowdwork (e.g., level of income or quality of output) are of less importance to them~-- thus the potential problems of collaborative work are not viewed as critically.

Looking at our sample, we identified a number of participants who worked for 2--4 hours per week and noted sentiments along the lines of: \textit{``I dislike 
[studies] that are oddly specific and do not give room for imagination to run''} (P18, Prolific).
Some casual workers may also be likely to have never encountered creative work on their platform, leading to statements such as the one by P85:
\textit{``Creative work does not exist on Prolific.''}%
%
%
%
\subsubsection{The Novelty Seeker}%
This worker loves variety and seeks out new and interesting tasks.
The novelty seeker works only occasionally on CS platforms, as repetitive tasks on the platform will quickly cause the worker to be bored.
The worker thinks creative work is fun and tries to avoid monotonous work that may be perceived as too repetitive by this worker.
The worker enjoys the unexpected and thus prefers working on creative tasks that the worker has never seen before.
Games and collaborative experiments are seen as exciting and memorable experiences in the view of this worker.
Unlike the professional worker, this worker is open to collaborative tasks.
The novelty seeker might produce low quality work if given too 
repetitive and monotonous tasks, or tasks that the worker has already performed before.

Among our sample, several participants mentioned testing new products, applications, or games for the sake of gaining new experiences. An example was mentioned by a worker (P67) from Prolific: \textit{``I enjoy undertaking more creative work or tasks where there is something different to do i.e. games, websites etc.''} The same worker hoped to see more creative work being offered on Prolific, as such work often entails new experiences.%
%
%
\subsubsection{The Self-Developer}%
The self-developer is primarily intrinsically motivated and seeks tasks that will make workers learn something or gain knowledge about themselves. The self-developer strives for continuous improvement.
This type of worker loves creative work, for a variety of reasons.
Creative tasks make the worker self-reflect and learn something about their own personality.
Debriefing is important for this worker, as the worker wants to see results.
Compensation is less important to this worker, but fair compensation still matters.

In our sample, various participants brought up self-development in conjunction with problem solving, e.g., \textit{``solving creative problems and tasks with other workers and myself because it helps exercise my brain and expand my learning horizons''} (P199, MTurk).
Some participants brought up the point 
that participating in research often leads the workers to discover things about themselves: \textit{``I think it is very rewarding for both workers and requesters in both finding out about themselves and furthering their research respectively''} (P107, Prolific).
Similarly, this type of worker enjoys being challenged, as \textit{``some studies raise interesting questions about [the worker's] own beliefs/attitudes''} (P13, Prolific).%
%
%
\subsubsection{The Pragmatic Worker}%
Regardless of working hours or other worker characteristics, the pragmatic worker holds the view that the worker's opinions do not matter. The requesters request, and the workers choose to work on tasks on their own volition, exercising their rights to decline a task they do not see worthy of doing.
Clear information about a task is key for this worker. This worker is often skeptical of creative work, as it often does not pay well enough, the task duration may be difficult to estimate, and the pay may not be worth the effort.
If the pragmatic worker chooses to work on creative tasks, the worker will answer with concentration and honesty, but may have doubts in their own ability to be creative.

In our analysis, we found the pragmatic worker to be \textit{``ambivalent''}
about creative work (P2, Prolific).
The pragmatic worker's mindset manifested 
in statements such as \textit{``Just state clearly the type of tasks that are part of the study beforehand. Then I can chose to do it or not''} (P89, Prolific), or the following response from P145, a worker with a 5-year working history on MTurk who worked up to 40~hours per week: \textit{``I don't have any preference for what kind of work is on mturk as long as the pay is fair and the requester is fair.''}
%
%
%
%
\subsection{Nonna\"{i}vet\'{e} of Crowd Workers}%
The existence of workers that are likely to have participated in academic studies has been documented in prior literature~\cite{chandler2013.pdf}.
Our study found evidence of worker nonna\"{i}vet\'{e} in regard to common creativity tests, as a subset of creative work on CS platforms.
Together with the quantitative data on prior exposure to creativity tests and creative tasks (Figure~\ref{fig:likertgraphs}), the past encounters raise a point to consider: is the data collected from these participants on creativity tests \todo{valid}?
Creativity tests that rely on an ``a-ha'' experience, such as Practical Insight Problems~\cite{Weisberg_Creativity.pdf}, are particularly affected by prior exposure.
If a relatively large portion of the sample has prior exposure potentially to the very same creativity test, this should be considered in the study design and participant screening.

And what crowdsourcing platform should be used for eliciting creative work and studying creativity? 
In studies in HCI, the ability and willingness of the crowd to participate in creative work seems to be largely unquestioned.
Our work found differences in how workers perceive creative work. Professional workers may have a negative attitude towards creative work, as it may take more time to complete and may be associated~-- in the view of the worker~-- with considerable uncertainty in regard to the rewards.
Our data indicates that casual workers on Prolific have been less exposed to creativity studies, show more interest in creative work, and may therefore be a better participant pool for creativity-oriented research.

\subsection{Actionable Insights}
From our analysis, we distill recommendations for requesters of creative work, independent of the crowdsourcing platform.%
%
\subsubsection{On Qualification Criteria}%
Literature on crowdsourcing recommends setting high qualification criteria for crowdsourcing studies as a measure to improve the quality of the responses. Matherly, for instance, suggested using an approval rate of 99\%~\cite{matherly_2018_ejm.pdf}.
For creativity studies, and especially creativity tests, setting high qualification criteria may be counter-productive~\cite{MTurkSamplingPractices_PrePrint_5-June-2019.pdf}. Limiting the pool of workers to professional workers in creativity studies will increase the probability of worker nonna\"{i}vet\'{e} and thus may negatively impact the study.
We caution the HCI community to
consider this trade-off when deciding on qualification criteria.
%
%
\subsubsection{On Rewards and Motivation}%
Another trade-off to consider is the reward for creative work.
It has previously been suggested that requesters should find a ``sweet spot'' of payment, as overpayment may attract spammers and lower the data quality~\cite{bohannon2011.pdf}. This general guideline, at first blush, may appear to make good sense.
In creative work, however, we found some types of workers are less motivated by extrinsic factors and may care more about opportunities for learning and entertainment than monetary compensation.
Requesters of creative work should think about how participants can be motivated, beyond monetary rewards. Requesters should follow up with results, as some workers want to know what their data is used for.~Self-developers,~in~particular, benefit from debriefing and post-task feedback.%

Fair compensation, 
however, should still be a priority in creative work, as highlighted by the general trend 
and pervasive complaints in our study about creative work being underpaid.
Connected to setting a price, requesters of creative work should be transparent and accurate about the time and effort required
to allow workers to make an informed decision for self-selecting to participate. Time traps, such as delays caused due to collaborative work with other crowd workers, should be avoided, especially if the participants include professional workers who prefer to work alone.%
%
%
\subsubsection{On Task Design}%
Providing clear task instructions and uncomplicated user interfaces is key to ensure requesters do not detract from the value of professional workers, who may be well adapted to provide creative input, but may shun creative work in general.
In this vein, informing workers about the goals of the task is also an important design component of a successful creativity study.

\rev{As for the type of creative work offered, self-developers and novelty seekers may be well aligned with the mini-c type of creative work.
The Big-C approach to studying creativity is not suitable for anonymous CS platforms, as traditionally, this type of creativity is studied through analysis of biographical profiles. One can argue that eminent contributions 
are unlikely to be elicited on microtask CS platforms. 
A Pro-C approach to creative work may also not be suitable for general-purpose CS platforms and better suited to other platforms that attract people with particular creativity-oriented skill sets.}%
%
%
%
%
\subsubsection{On Validity Issues}%
We caution researchers using crowdsourcing platforms for creativity studies to carefully reflect on the aims of their study, the choice of platform, and the validity of the data.
The validity may be affected not only by prior exposure to tasks, but also the attitude of workers towards a creative task.
Some workers, and according to our analysis especially professional MTurk workers, may be disinclined to participate in creative work.
Participant screening and convening participant pools specific for creative work may be two solutions in this regard.%
%
%
\subsection{Future of Creativity Research using {Paid CS Platforms}}%
%
\subsubsection{Alone or Together?}%
Collaborative crowd work has been gaining attention in the form of  different collaborative workflows and team experiments with workers recruited from crowdsourcing platforms (e.g., \cite{FLASHTEAMS,FLASHORGS}).
Further, studies of collaborative creative processes dominate the field of creativity research in HCI 
\cite{p1235-frich.pdf}.
Our study found evidence that workers strongly prefer to work alone. In particular professional crowd workers are prone to avoid collaborative tasks due to their precarious nature.
Given that professional workers were found to complete the majority share of tasks on \todo{paid crowdsourcing platforms}~\cite{p135-difallah.pdf,Caution-MTurk-Workers-Ahead-Fines-Doubled.pdf,PIIS1364661317301316.pdf}, creativity researchers need to be aware that a large part of this very same group of the professional workers tends to have a negative attitude towards, or simply does not care about, creative work.%
%
\subsubsection{Should there be more Creative Tasks?}%
From the viewpoint of many workers, creative work is seen as both beneficial and wanted. The variety it offers is something valuable, but there are considerations to be taken into account.
Novelty seekers and self-developers look for variety in their tasks, but pragmatic and professional workers prefer to simply do as instructed.
Even though in theory the scope of creativity offers limitless possibilities for tests and tasks, in reality repetition still occurs.
Different worker profiles can react differently to approaches taken to alleviate this issue.
For instance, one could devise a work flow that lets workers select a creativity test that they have never completed before.
The benefits of such a strategy, though, may be lost on professional workers who may prefer known tasks, as such tasks can be completed more quickly.
This should be taken into consideration when designing creativity studies on crowdsourcing platforms.%
%
%
\subsubsection{Purposes and Outcomes of Creative Work}%
With creative work, issues with idea ownership and data use are likely to emerge~– even if currently ignored by the majority of requesters.
The Federal Trade Commission (FTC), for instance, is becoming increasingly interested in data security, implying that any crowdsourcing enterprise that collects and redistributes data must protect the data and inform all stakeholders of what information the enterprise collects and how it will use the information~\cite{WolfsonLookBeforeLeap}.
Similarly, if enterprises end up using creative ideas contributed by the crowd, what does this mean for patentable inventions?
The potential in crowd-powered creativity is, it seems, well-matched by the anticipated legal and ethical concerns. 
The reasonable way forward is negotiation. We need to get workers who are interested in creative work~-- in donating their potentially valuable ideas for pennies on the dollar~-- to the same table with requesters, platform operators and 
policymakers to strike trade-offs and find common ground. 
The current model of crowdsourcing platforms hardly supports this.%
%
%
\subsubsection{A Design Space for Creative Work on {Paid CS Platforms}}%
Prior literature suggested tailored platforms for human-centered experiments~\cite{Gadiraju2-1wgg9voexzjhw3.pdf}. In essence, Prolific already aims to be such a platform for behavioral research. 
Future tailored platforms could particularly target and cater to participants who enjoy \todo{creative microtask work}.
Alternatively, recruitment tools interfacing with current crowdsourcing platforms~-- similar to TurkPrime~\cite{TurkPrime}~-- could be created to help with cultivating a participant pool interested in creativity studies. 
To this end, a more complete understanding of the design space for creative work on crowdsourcing platforms is needed, for which our study provides the first piece: the viewpoint of the worker.%
%
\subsection{Study Limitations}%
The responses to our survey 
need to be considered with care, as 
workers aim to maximise their income by completing tasks as quickly as possible~\cite{p218-gadiraju.pdf}.
It is therefore possible that some workers
did not 
complete the questionnaire in a deliberate manner.
Further, workers self-select to participate and results may therefore be biased~\cite{couper2000.pdf}. For instance, our results may be biased towards workers who specialize in taking surveys and may not 
represent the whole population of crowd workers on the respective platform.
Further, crowdsourcing contributes to the livelihood of many 
workers. Thus, there is a possibility some participants may have purposefully shaped their answers to live up to our expectations and not curb future work opportunities.


Our qualification criteria on MTurk excluded workers who have completed few HITs and may therefore be biased towards experienced workers who have been exposed to measures commonly used in creativity-oriented studies.
However, recruiting participants by a combination of the number of HITs previously approved and the approval rate is a common practice on MTurk~\cite{peer2013.pdf}.
Given that the Prolific workers were not complete novices and well qualified, and there are no commonly-used recommendations for qualification criteria available for Prolific, we believe our selection of participants 
is justified.%
%
%
\section{Conclusions}%
In this paper, we provided a timely look into the workers' perspective as an understudied area in crowd-powered creative work.
Our analysis revealed clear differences between the workers available on two commonly used platforms, MTurk and Prolific.
An in-depth qualitative analysis of the workers' responses to an online questionnaire revealed reasons why certain workers like and other workers resent creative work, such as learning while working versus the precarious nature of creative work and its rewards.
Further, we discussed several worker archetypes derived from our analysis and provided takeaways for requesters.
Certainly, much about using \todo{paid crowdsourcing platforms} for studying creativity or eliciting creative input remains to be investigated. 
Our findings provide a first starting point for a meaningful discussion about the use of crowdsourcing platforms in creativity-oriented research.%
%
%
%
\section{Acknowledgments}%
This work was in part funded by the Academy of Finland (Grants 313224-STOP, 320089-SENSATE, 316253-SENSATE and 318927-6Genesis Flagship).


\balance{}

\bibliographystyle{SIGCHI-Reference-Format}
\bibliography{paper}

\end{document}